\documentclass[10pt,journal,compsoc]{IEEEtran}
\usepackage[nocompress]{cite}
\usepackage{url}
\usepackage{amsmath,amssymb,amsfonts}
\usepackage{multirow, threeparttable}
\usepackage{lineno}
\usepackage{graphicx}
\usepackage{stfloats}
\usepackage{algorithm, algorithmic}
\usepackage{color}
\usepackage[caption=false,font=footnotesize,labelfont=sf,textfont=sf]{subfig}

\begin{document}
\title{Hierarchical Deep Reinforcement Learning for VWAP Strategy Optimization}

\author{Xiaodong~Li,
        Pangjing~Wu,
        Chenxin~Zou,
        and~Qing~Li,~\IEEEmembership{IEEE Fellow}
\IEEEcompsocitemizethanks{\IEEEcompsocthanksitem X. Li, P. Wu and C. Zou are with the College
of Computer and Information, Hohai University, Nanjing, China, 211100.\protect\\
E-mail: xiaodong.li@hhu.edu.cn
\IEEEcompsocthanksitem Q. Li is with Department of Computing, The Hong Kong Polytechnic University, Hong Kong, China, 999077.}
\thanks{Manuscript received MM DD, YY; revised MM DD, YY.}
\thanks{Corresponding author: Xiaodong Li.}}

\markboth{Transactions on Big Data,~Vol.~XX, No.~XX, MM~YY}%
{Li \MakeLowercase{\textit{et al.}}: Hierarchical Deep Reinforcement Learning for VWAP Strategy Optimization}

\IEEEtitleabstractindextext{%
\begin{abstract}
Designing an intelligent volume-weighted average price (VWAP) strategy is a critical concern for brokers, since traditional rule-based strategies are relatively static that cannot achieve a lower transaction cost in a dynamic market. Many studies have tried to minimize the cost via reinforcement learning, but there are bottlenecks in improvement, especially for long-duration strategies such as the VWAP strategy. To address this issue, we propose a deep learning and hierarchical reinforcement learning jointed architecture termed Macro-Meta-Micro Trader (M3T) to capture market patterns and execute orders from different temporal scales. The Macro Trader first allocates a parent order into tranches based on volume profiles as the traditional VWAP strategy does, but a long short-term memory neural network is used to improve the forecasting accuracy. Then the Meta Trader selects a short-term subgoal appropriate to instant liquidity within each tranche to form a mini-tranche. The Micro Trader consequently extracts the instant market state and fulfils the subgoal with the lowest transaction cost. Our experiments over stocks listed on the Shanghai stock exchange demonstrate that our approach outperforms baselines in terms of VWAP slippage, with an average cost saving of 1.16 base points compared to the optimal baseline.
\end{abstract}

}

\maketitle
\IEEEdisplaynontitleabstractindextext

\IEEEraisesectionheading{\section{Introduction}\label{sec:introduction}}
\IEEEPARstart{W}{hen} executing a large order (\emph{a.k.a.} a parent order), brokers may encounter a large negative slippage caused by market impacts. The large order makes the market price fall (or rise) suddenly, resulting in the actual trading price being far from expected, which increases the transaction costs. An effective way to solve this problem, known as algorithmic trading, is to split the parent order into several small orders (\emph{a.k.a.} child orders) over time and execute these orders according to a preset rule.

In 1988, Berkowitz~\emph{et al.}~\cite{berkowitz1988total} suggested using the difference between market volume-weighted average price (VWAP) and orders' VWAP as a metric for transaction costs, \emph{a.k.a.} VWAP slippage. They split a parent order based on the market liquidity to reduce the VWAP slippage, known as the VWAP strategy. Since it needed few assumptions about market microstructures~\cite{lin2020an}, brokers widely adopt it in order execution. However, this static rule-based strategy cannot obtain a lower transaction cost in a dynamic market. Researchers began to improve the VWAP strategy from two aspects. One was to improve the estimation of volume profiles~\cite{konishi2002optimal}\cite{bialkowski2008improving}. The other was to adjust the trading strategy according to recent historical price and volume information~\cite{kakade2004competitive}\cite{almgren2006bayesian}. Nevertheless, these approaches relied on strict modeling and assumptions about the market's microstructures, and the unpredictable market limited their efficiency~\cite{lin2020an}.

Since stock trading is essentially a decision process consistent with the problem that reinforcement learning (RL) aims to solve, many researchers began to optimize order execution strategies via RL models. Nevmyvaka~\emph{et al.}~\cite{nevmyvaka2006reinforcement} first revealed the efficiency of RL in optimizing transaction costs by a large-scale empirical application. Then many researchers began to leverage the RL models to optimize the order execution strategies~\cite{lin2020an}\cite{ning2018double}. However, these studies mainly focused on strategies executed for a few minutes, as shown in Table~\ref{tab:summary}. They ignored the challenging strategies that took longer time to execute, such as the VWAP strategy. As many studies~\cite{niederhoffer1966market}\cite{hasbrouck1999dynamics} demonstrated that there was a significant daily periodicity in trading intensity, the fluctuating intensity and complicated market patterns increased the complexity of optimal decision making. It is vital to design an RL-based trader that can capture variations of price and liquidity throughout the day as well as micro changes within a few seconds on the limit order books (LOBs) without making strict assumptions.

In addition, although recent representative research~\cite{mnih2015human} revealed that deep Q-learning network (DQN) outperformed human players in most Atari games, it failed to learn an effective policy in \textit{Montezuma's Revenge}, where there were various goals and sparse rewards. A practical approach to address this problem was known as hierarchical reinforcement learning (HRL), which solved this problem by establishing temporal abstractions for the environment and decomposing the decision process into several simpler subgoals, showing good adaptability and efficiency in addressing complex RL tasks~\cite{kulkarni2016hierarchical}.

By incorporating the idea of hierarchical decision, we propose a deep learning (DL) and HRL jointed architecture termed Macro-Meta-Micro Trader (M3T) to optimize the VWAP strategy. The M3T captures market patterns and executes orders from different temporal scales by three hierarchical traders: Macro Trader, Meta Trader, and Micro Trader. It optimizes the VWAP strategy by improving trading volume estimation, parent order allocation, subgoal selection, and child order execution. Firstly, the Macro Trader adopts a long short-term memory (LSTM) network~\cite{hochreiter1997long} to estimate future volume profiles and allocates a parent order to tranches accordingly. Secondly, the Meta Trader selects a subgoal appropriate to instant liquidity to form a mini-tranche. Finally, the Micro Trader extracts the market state by a multi-headed self-attention encoder and executes orders to fulfils the subgoal and reduce transaction costs. To verify the efficiency of our approach, we perform extensive experiments over eight stocks listed on the Shanghai stock exchange (SSE). The experimental results demonstrate that our M3T outperforms baselines in terms of VWAP slippage on all stocks, with an average cost saving of 1.16 base points compared to the optimal baseline, \emph{i.e.}, double DQN (DDQN)~\cite{van2016deep}. Further analysis also demonstrates that each trader in M3T can effectively reduce the transaction costs.

Our main contributions are summarized as follows:
\begin{enumerate}
	\item We propose a DL and HRL jointed architecture termed M3T, which optimizes the VWAP strategy from multiple layers of trading volume estimation, parent order allocation, subgoal selection, and child order execution. The optimal VWAP strategy is obtained by optimizing each layer hierarchically in the M3T. To our best knowledge, it is the first study incorporating RL into VWAP strategy optimization.
	\item We extend Markov decision process (MDP) to hierarchical Markov decision process (HMDP), introducing hierarchical control into the decision process. We also consider microstructures of the Chinese stock market that are quite different from previous studies in trading simulation and the design of HMDP, such as minimal trade size.
	\item We discuss the effectiveness of different DL and RL models on the VWAP strategy optimization, and the experimental results verify the efficiency of the M3T.
	
\end{enumerate}

The rest of this paper is organized as follows. Section~\ref{sec:rw} discusses the related studies on algorithmic trading and HRL. Section~\ref{sec:pf} introduces the VWAP strategy and proposes HMDP for M3T optimization. Section~\ref{sec:m3t} proposes our M3T architecture. Section~\ref{sec:exp} describes the experimental setup and reports the experimental results and related discussions. Section~\ref{sec:con} provides our conclusions.

\section{Related Work}
\label{sec:rw}
\subsection{Algorithmic Trading}
Bertsimas and Lo~\cite{bertsimas1998optimal} were the first to use dynamic programming to find the explicit closed-form solution of order execution. Almgren and Chriss~\cite{almgren2001optimal} proposed the Almgren-Chriss model, extending Bertsimas and Lo's method by adding cost evaluation, price impact functions, and risk aversion parameters. Huberman and Stanzl~\cite{huberman2005optimal} considered traders' appetite for liquidity risk in order execution. However, these studies make strong assumptions about underlying price movements or distributions, which are difficult to apply to real stock markets. Berkowitz~\emph{et al.}~\cite{berkowitz1988total} suggested using VWAP slippage as the metric of the transaction cost. They proposed the VWAP strategy that split a parent order based on the market liquidity to reduce the VWAP slippage. However, these static strategies could not obtain optimal costs in a dynamic market. There were two main approaches to enhancing the VWAP strategy. The one was to improve the forecasting of volume profiles. Konishi~\cite{konishi2002optimal} modeled the fluctuation pattern of price and volume as Brownian movement. Bialkowski~\emph{et al.}~\cite{bialkowski2008improving} proposed a volume forecasting model combining market features and stock features. The other was to improve the strategy using recent price and volume information. Kakade~\emph{et al.}~\cite{kakade2004competitive} introduced the competitive ratio into the VWAP strategy to reflect the market's liquidity and dynamically adjusted child orders. Almgren and Lorenz\cite{almgren2006bayesian} minimized the transaction costs via Bayesian estimators of intraday price shifts. However, these approaches mainly relied on the assumptions of markets' microstructures, such as market dynamics, order types, trading rules, etc. Markets' unpredictable patterns limited their capabilities.

As reinforcement learning brought a new way to find the optimal order execution strategy by automatically learning the underlying dynamic of markets from massive data, many studies began to leverage RL methods to optimize trading strategies dynamically. Table~\ref{tab:summary} summarizes the existing studies on RL-based algorithmic trading strategies for sale side. The current studies mainly focus on orders with short execution times and small order sizes, which still has a big gap with actual trading task requirements. Nevmyvaka~\emph{et al.}~\cite{nevmyvaka2006reinforcement} proposed the first large-scale empirical application of optimizing order execution via RL. They minimized the implementation shortfall (IS) for a buy-side (or sell-side) within a short discrete period. Hendricks and Wilcox~\cite{hendricks2014reinforcement} optimized IS by using Q-learning to revise the volume derived from the Almgren-Chriss model. Recently, the advent of the DQN~\cite{mnih2015human} has provided a powerful representation ability for RL agents. Ning~\emph{et al.}~\cite{ning2018double} used DQN directly to train trading agents based on high-dimensional LOB data. Ye~\emph{et al.}~\cite{ye2020optimal} used convolution neural networks to extract spatial information of LOBs and minimized the transaction costs by deep deterministic policy gradient. Lin and Beling~\cite{lin2020an} leveraged LSTM networks to extract features from high-dimensional market information as agents' observations and optimized the IS with the proximal policy optimization method. Nevertheless, the existing studies mainly focused on optimizing short-duration order execution and paid less attention to others. 

\begin{table}[t]
	\caption{Existing Studies of RL in Algorithmic Trading.\label{tab:summary}}
	\begin{threeparttable}
		\begin{tabular*}{\hsize}{@{\extracolsep{\fill}}|l|r|r|}
			\hline
			\textbf{Author} &  \textbf{Execution Time} & \textbf{Total Order Size} \\
			\hline
			Nevmyvaka~\emph{et al.}~\cite{nevmyvaka2006reinforcement} & 2$\sim$8 min & 5K; 10K \\
			Hendricks and Wilcox~\cite{hendricks2014reinforcement}	& $20\sim60$ min	& 10K; 1M\\
			Shen~\emph{et al.}~\cite{shen2014risk}	&	10 min	&	20K	\\
			Ning~\emph{et al.}~\cite{ning2018double}	&	60 min	&	2K	\\
			Ye~\emph{et al.}~\cite{ye2020optimal}	&	2 min	&	5K; 10K	\\
			Lin and Beling~\cite{lin2020an}	&	2 min	&	$300\sim7$K	\\
			Fang~\emph{et al.}~\cite{fang2021universal}	&	30 min	&	$<1$K	\\
			Wang~\emph{et al.}~\cite{wang2021commission}	&	20 min	&	$10$K$\sim40$K	\\
			\textbf{Ours}	&	240 min	&	$100K\sim160K$\tnote{*} \\
			\hline
		\end{tabular*}
		\begin{tablenotes}
			\item[*] The total order size of our method is proportional to the given minimum trading unit. Increasing the minimum unit allows our method to be adapted to larger orders.
		\end{tablenotes}
	\end{threeparttable}
\end{table}

Besides, some studies on portfolio management have also noticed the transaction costs caused by the adjustment of portfolio position in recent years. Fang~\emph{et al.}~\cite{fang2021universal} suggested optimizing algorithmic trading strategies based on knowledge distillation. They used a teacher network with future information to guide a student network with historical data to learn the optimal strategy. Wang~\emph{et al.}~\cite{wang2021commission} proposed a two-layer hierarchical RL model, separating the portfolio management and order execution. The upper agent determined the portfolio position ratio, while the lower agent adjusted the position and minimized transaction costs. However, the position adjustment window was short. These methods' effectiveness was only empirically verified on a short time scale. Therefore, designing an RL-based trader sensitive to both micro and macro changes in price and liquidity throughout the day was significant.

\subsection{Hierarchical Reinforcement Learning}
Sparse rewards and long execution times presented tremendous challenges for training RL agents. An effective way to address the challenges was by establishing temporal abstractions to decompose decision processes into simpler subgoals. Dayan and Hinton~\cite{dayan1993feudal} first proposed the idea of multi-level control by dividing the task into three levels: super-manager, manager, and sub-manager. All the levels obeyed two principles: 1) reward hiding, the agent only needed to fulfil the task given by the immediate superior agent rather than considering the completion of the overall task; 2) information hiding, the sub-manager did not need to know the goal issued by the super-manager, and the super-manager did not need to know how the sub-manager achieved the goal. This work provided a prototype for subsequent studies, guiding the design of HRL structures.

After that, Sutton~\emph{et al.}~\cite{sutton1999between} combined the idea of temporal abstraction over action set with MDP and SMDP. In their option framework, the upper agent selects an original action or a combination of original actions as an option in each step, while the lower agent defines a policy over actions for each option.
Sorg and Singh~\cite{sorg2010linear} optimized the options by combining existing options. Szepesvari~\emph{et al.}~\cite{szepesvari2014universal} optimized the options by using different reward functions.
Bacon~\emph{et al.}~\cite{bacon2017option} combined the option framework with the actor-critic structure and directly adopted the policy gradient algorithm to synchronously optimize multi-level control.
Besides, some studies have used mutual information rewards to drive low-level agents to actively explore the dynamics of actions and environments without environmental rewards~\cite{mohamed2015variational, eysenbach2018diversity, goyal2019reinforcement}.

From the perspective of state and goal abstractions, Schaul~\emph{et al.}~\cite{schaul2015universal} proposed a universal value function for multi-goal optimization problems, which considered goals along with states. Kulkarni~\emph{et al.}~\cite{kulkarni2016hierarchical} proposed a temporal abstract HRL, which divided the overall goal into several subgoals and learned an effective global strategy over each subgoal. It showed good adaptability and efficiency in addressing complex RL tasks. Nachum~\emph{et al.}~\cite{nachum2018data} solved the non-stationary problem between the upper and lower agents in off-policy HRL optimization. Besides, some HRL studies focused on how to learn effective policy based on the idea of hindsight in an environment with sparse rewards. Andrychowicz~\emph{et al.}~\cite{andrychowicz2017hindsight} proposed the hindsight experience replay to select efficient goals under sparse rewards. The effective policy was learned on the failed transition by replacing the unreachable goal with the last reached state. The transition would then receive a positive reward. Since the shift of lower-level policy would affect the dynamic of the upper level, non-stationary problems occurred that previously achievable goals could not be achieved. Levy~\emph{et al.}~\cite{levy2019learning} assumed that the policy at the lower level was already optimal, so changes in the policy at the lower level would not affect the policy at the upper level.

\section{Problem Formulation}
\label{sec:pf}
\subsection{Preliminary}
Many studies~\cite{niederhoffer1966market}\cite{hasbrouck1999dynamics} have demonstrated a significant daily periodicity in trading intensity (as shown in Fig.~\ref{fig:vp}), which may affect order execution. When the market lacks liquidity, issuing a large order can lead to price deviation from the expected or even failure of order execution, thus increasing transaction costs. An effective way to reduce the transaction costs caused by liquidity risks is leveraging algorithmic trading, which allocates large orders into several small tranches according to pre-programmed trading strategies in order to minimize the impact on the market.

The VWAP strategy is a typical impact-driven algorithmic trading strategy focusing on predicting intra-day volume distribution instead of directly modeling the transaction costs. It is suitable for trading orders with large sizes and long execution times. In practice, it allocates a large order (the order's size is generally smaller than daily market liquidity or will be split into multiple days) into tranches according to the market's volume profiles. Each tranche's volume is consistent with the market liquidity. The expected average trading price is the VWAP, \emph{i.e.},
\begin{equation}
	\text{VWAP}=\frac{\sum_{t=1}^{T}{p_t \cdot v_t}}{T},
\end{equation}
where $v_t$ is the size of order issued at time $t$. The transaction costs of the VWAP strategy are the relative difference between the order's VWAP $p_{o}$ and the market's VWAP $p_{m}$, known as VWAP slippage, \emph{i.e.},
\begin{equation}
	\text{VWAP slippage}=\text{side}\cdot\frac{p_{o}-p_{m}}{p_{m}},
	\label{eq:slippage}
\end{equation}
where $\text{side}$ denotes the buying/selling direction, \emph{i.e.}, $-1$ for the buyer and $+1$ for the seller, so that a large positive VWAP slippage indicates a lower transaction cost.

However, the traditional VWAP strategy can only split parent orders according to markets' liquidity and executes orders according to a preset rule, such as the time weighted average price (TWAP) strategy. It cannot actively decide the timing and sizes of trading according to market conditions. There is still room to reduce transaction costs. Therefore, we propose a hierarchical RL architecture termed M3T to improve the VWAP strategy's parent order allocation and child order execution, which are described in the remainder of this section.

\subsection{Parent Order Allocation}
\begin{figure}[htb]
	\centering
	\includegraphics[width=\linewidth]{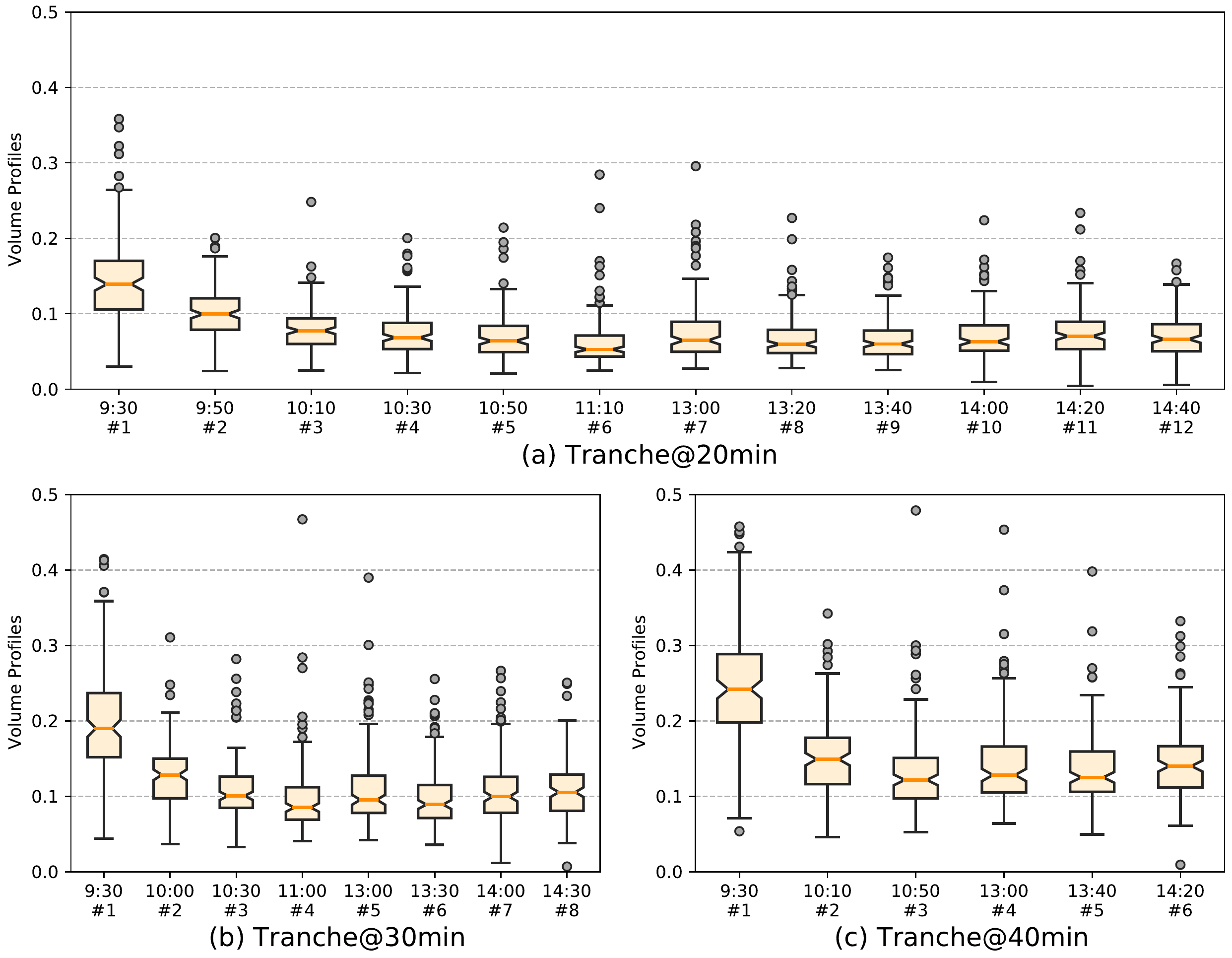}
	\caption{Historical volume profile distribution of Shanghai Pudong Development Bank (600000).
		\label{fig:vp}}
\end{figure}

Before executing a parent order, the VWAP strategy allocates the parent order into several tranches according to the volume profiles, a market liquidity metric represented by the proportion of each tranche's volume to the day's volume. The granularity is an adjustable parameter in the calculation of volume profile, affecting liquidity estimation accuracy, as shown in Fig.~\ref{fig:vp}. The fine-grained volume profiles have more outliers and noise, while the coarse-grained volume profiles eliminate daily periodicity and present a uniform distribution. Considering that the trading time of the Chinese stock market is 4 hours and the fine-grained volume profiles may contain noise, we divide a full-day execution order $\mathcal{T}$ into eight 30-minute tranches as Fig.~\ref{fig:vp} (b). In other words, on day $t$, the VWAP strategy first divides the last $n$ days' trading volume into eight 30-minute tranches; then it estimates the volume profiles $\widehat{vp}_t^{(i)}$ of tranche $\#i$ based on its past $n$ days' volume profiles $X_t^{\left(i\right)} \overset{.}{=} \left<vp_{t-n}^{(i)}, vp_{t-n+1}^{(i)}, \ldots, vp_{t-1}^{(i)} \right>$, where the volume profiles are estimated by moving average (MA) method in traditional VWAP strategy; finally it allocates the parent order into tranches according to the estimated volume profiles, and each tranche's order size is $\mathcal{T}_t^{\left(i\right)}=\mathcal{T}\times{\widehat{vp}}_t^{\left(i\right)}$.

\subsection{Trading Simulation}
Based on the microstructure of the Chinese stock market, we build a trading simulation system for order execution. In this system, an RL agent can only issue one child order at any time. The old order will be automatically canceled when a new order is issued, and the new order will enter different order queues according to its price. Take a sell order as an example:
\begin{itemize}
	\item If its price is on the buy-side, the order goes to the front of the queue and trades with buyers directly.
	\item If its price is on the sell-side, the order waits at the end of the queue. The position of the order is updated one step forward when recording a trade whose price is greater than or equal to the order.
\end{itemize}

However, it is impossible to correctly estimate the actual length of the queue based on the historical LOB and trade data, because we do not know how many orders are canceled or edited between each snapshot. Simply ignoring the length of the queue may make the agent too optimistic about the market liquidity. Therefore, we introduce an empirical constant $\iota$ (which is 3 in our experiments) as the estimator of the actual queue length. When the order reaches the head of the queue, it trades with all the quotes whose price is greater than or equal to the order price.

\subsection{Hierarchical Markov Decision Process}
MDP provides a mathematical model for training RL agents. In this paper, we extend traditional MDP to HMDP in order to facilitate training M3T. The HMDP consists of extrinsic state $\widetilde{\mathcal{S}}$, intrinsic state $\mathcal{S}$, subgoal $\mathcal{G}$, action $\mathcal{A}$, extrinsic reward $\widetilde{\mathcal{R}}$, intrinsic reward $\mathcal{R}$, and the state transition probability $\mathcal{F}$. Each element is defined as follows.

\noindent\textbf{Extrinsic State:} $\widetilde{\mathcal{S}}$ is the state observed by the Meta Trader. It consists of the market's liquidity, tranche execution progress, and overall market trading volume during the last subgoal execution. 

\noindent\textbf{Intrinsic State:} $\mathcal{S}$ is the state observed by the Micro Trader. It consists of LOBs, subgoal $g$, and subgoal execution progress, where the LOBs consist of the last 20 LOB snapshots (which are padded with 0 if the number of snapshots is less than 20); the subgoal $g$ is the one-hot embedding of the subgoal given by Meta Trader; the subgoal execution progress consists of time progress and filling ratio of $g$.

\noindent\textbf{Subgoal:} The action executed by the Meta Trader is named \textit{subgoal} $g=(T_g, V_g)$, which consists of the maximum execution step and order size that the Micro Trader should execute. The subgoal space $\mathcal{G}$ is preset based on the TWAP strategy. Considering a 30-minute tranche containing 600 discrete time steps, it must fulfil about 16\% of the orders for every 100 steps according to the TWAP strategy. We fluctuate the steps and the size by 20\% and obtain nine subgoals to adapt to the different market liquidity. These subgoals are shown in Eq.~\ref{eq:subgoal} and numbered from $\#1$ to $\#9$ line by line.

\begin{equation}
	\label{eq:subgoal}
	\mathcal{G}\!=\!
	\left[
	\begin{matrix}
		1\#(80,0.13)& 2\#(80,0.16)& 3\#(80,0.19)\\
		4\#(100,0.13)& 5\#(100,0.16)& 6\#(100,0.19)\\
		7\#(120,0.13)& 8\#(120,0.16)& 9\#(120,0.19)\\
	\end{matrix}
	\right].
\end{equation}

\noindent\textbf{Action:} In each step, the agent can issue a limit order or wait for the order to be fully filled. The alternative price of the limit order is current \textit{bid1} or \textit{ask1} price on the LOB, and the order's size is fixed to one lot (\emph{i.e.}, 100 shares), which is the minimum trading unit in the Chinese stock market. Besides, when the Micro Trader finally fails to fulfil the subgoal, the remainder is liquidated as a market order.

\noindent\textbf{Extrinsic Reward:} According to Sutton~\emph{et al.}~\cite{sutton2018reinforcement}, rewards should directly reflect the goal rather than how to achieve it. Therefore, we directly use the VWAP slippage (Eq.~\ref{eq:slippage}) as the extrinsic reward $\widetilde{r}\in\widetilde{\mathcal{R}}$. The Meta Trader will receive a VWAP slippage when the subgoal is fulfilled and 0 otherwise. Besides, if the Micro Trader finally fails the subgoal, the environment will impose a corresponding penalty according to order size and market liquidity. Specifically, the Meta Trader is considered to give an inappropriate subgoal if the minimum step of fulfilling the subgoal exceeds 30\% of the maximum step specified by the subgoal or the order size of the subgoal exceeds 5\% of the market liquidity. Otherwise, the Micro Trader will be punished.

\noindent\textbf{Intrinsic Reward:} The intrinsic reward $r\in\mathcal{R}$ also feeds back with a VWAP slippage (Eq.~\ref{eq:slippage}). It gives 0 at each step and returns the VWAP slippage when the subgoal is terminated, called sparse reward. Meanwhile, considering that sparse rewards may harm the performance of general RL models, we propose a dense reward, which feeds back with the VWAP slippage of current filled orders or 0 for otherwise. These two rewards are used in different models. Besides, in reality, brokers must fill clients' orders within a given time. The Meta Trader and the Micro Trader will receive a penalty of $-99$ if the order is not filled at the end of each subgoal. In addition, we ignore the commission fees because the commission fees in the Chinese stock market are fixed and do not affect the evaluation of transaction costs.

\noindent\textbf{State Transition:} Since our environment is a simulation of the real Chinese stock market based on historical data, we cannot correctly estimate the impact of our agent's decision on the market. However, a liquid market will rebound to balance after impacting~\cite{lin2020an}, because the child order issued by our agents is tiny. Therefore, we assume that $\mathcal{F}$ is consistent with the history, just as previous studies did~\cite{nevmyvaka2006reinforcement}\cite{ning2018double}\cite{lin2020an}\cite{ye2020optimal}.

\section{Macro-Meta-Micro Trader}
\label{sec:m3t}
\label{sec:model}
\begin{figure}[htb]
	\centering
	\includegraphics[width=0.95\linewidth]{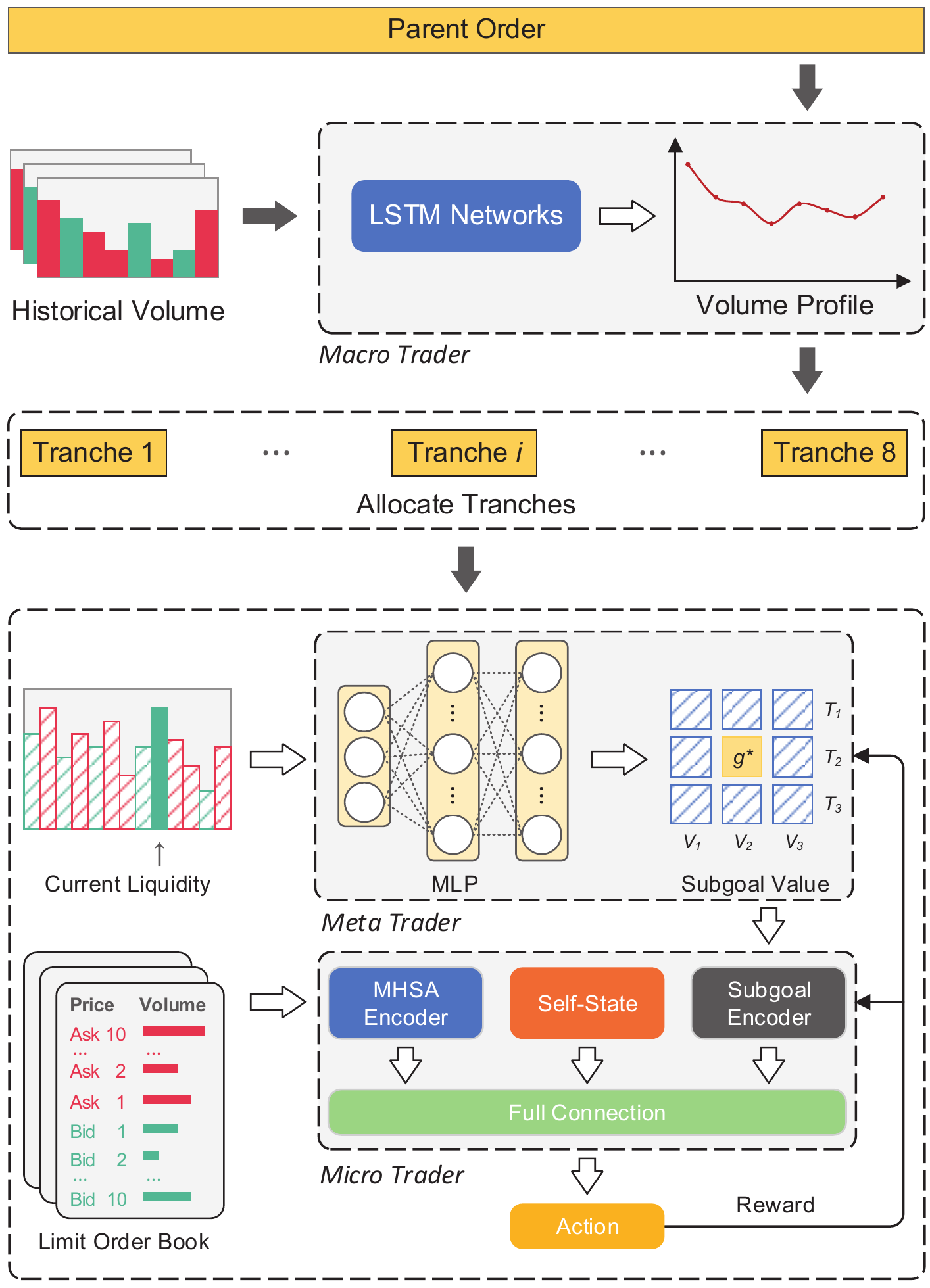}
	\caption{The scheme of the M3T for VWAP strategy optimization. A parent order is firstly allocated into tranches based on volume profiles predicted by Macro Trader. Within each tranche, Meta Trader then selects a subgoal appropriate to the current liquidity for Micro Trader, which executes orders based on raw LOB data to fulfil the subgoal. Finally, Meta Trader and Micro Trader optimize their policies according to the feedback rewards.}
	\label{fig:process}
\end{figure}

In this section, we propose our M3T architecture, which consists of Macro Trader, Meta Trader, and Micro Trader, as illustrated in Fig.~\ref{fig:process}.

\subsection{Macro Trader}
\begin{figure}[!htb]
	\centering\includegraphics[width=0.9\linewidth]{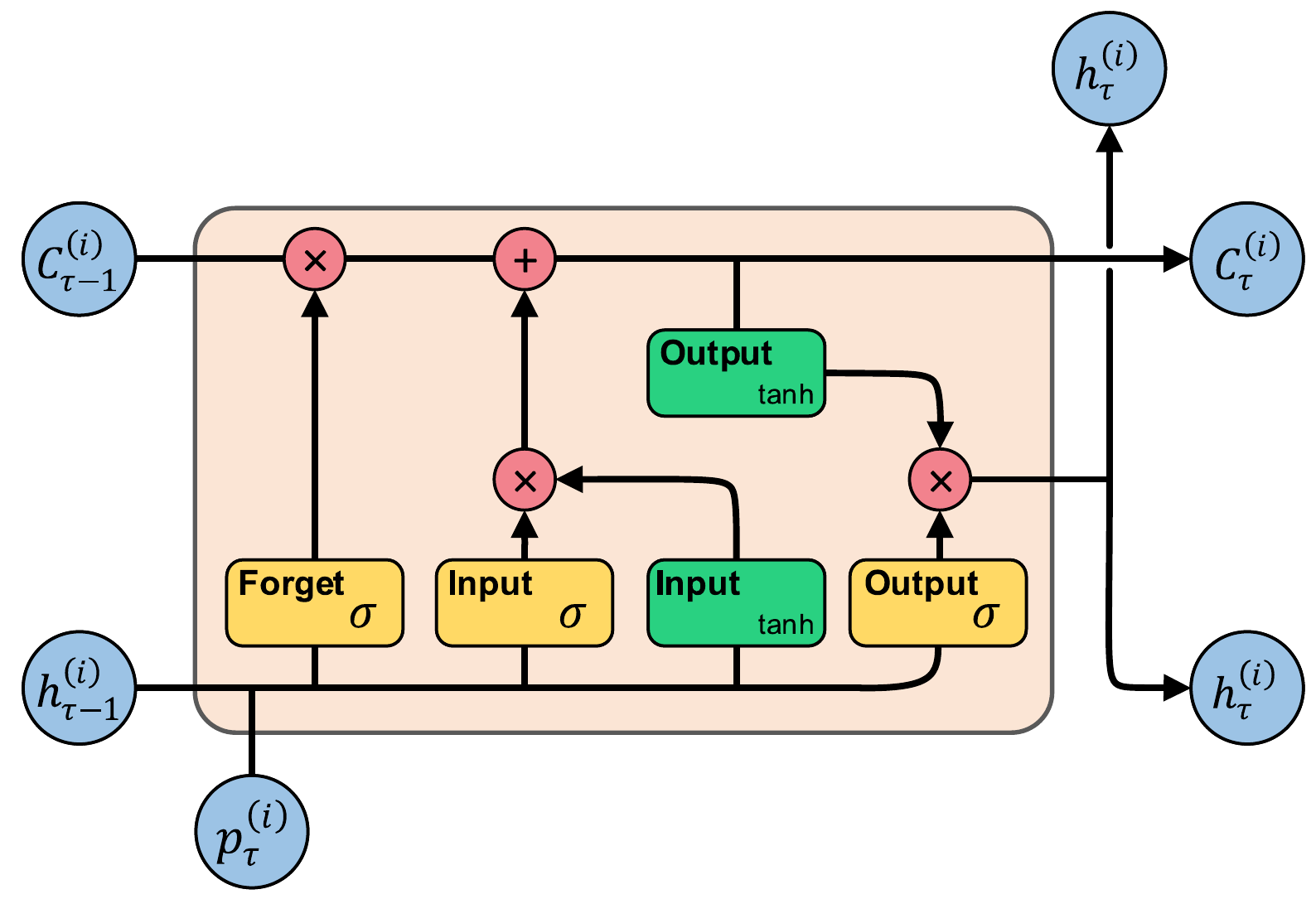}
	\caption{An example of using LSTM to forecast the $i$th tranche's volume profile $p_t^{(i)}$ in day $t$ based on last $n$ days' volume profiles. At time step $\tau \in [t-n, t-1]$, the inputs of the LSTM unit are current volume profile $vp_{\tau}^{(i)}$, last hidden state $h_{\tau-1}^{(i)}$ and cell state $C_{\tau-1}^{(i)}$. The LSTM first decides to discard which information from the cell state by the forget gate. Then, the input gate obtains temporal information from inputs to form a candidate cell state. Finally, the output gate decides to output which parts of the cell state as the temporal representation of volume profiles.}
	\label{fig:lstm}
\end{figure}

To reduce parent orders' impact on the market, we leverage Macro Trader to forecast volume profiles and allocate parent orders into tranches. It uses LSTM~\cite{hochreiter1997long} as the backbone model to learn the temporal pattern of volume distribution. The LSTM network is a recurrent neural network (RNN) widely used to capture temporal patterns. Unlike traditional RNN~\cite{hopfield1982neural} that only contains recurrent information flow, an LSTM unit consists of several cells, \emph{i.e.,} a forget gate, an input gate, and an output gate, to control the input and output of temporal information, as shown in Fig.~\ref{fig:lstm}. These gate structures overcome the vanishing and explosion of gradient encountered by the RNN in extracting temporal representations.

Specifically, the Macro Trader first uses the LSTM network to extract the temporal pattern of the last $n$ days' historical volume profiles tranche-by-tranche. Then a full connection (FC) layer after the LSTM network maps the hidden state output by the LSTM to estimate the volume profiles $\left<vp_t^{(1)}, vp_t^{(2)}, \ldots, vp_t^{(8)}\right>$ of day $t$. Finally, according to the estimated volume profiles, it allocates a full-day executed parent order to eight 30-minute tranches.

\subsection{Meta Trader}
Within each tranche, the market liquidity still fluctuates slightly under events' impact, which harms the performance of existing RL models. Further splitting the tranche is susceptible to extreme short-term liquidity, enlarging the bias of volume profile estimation. Therefore, we establish Meta Trader to bridge the Macro Trader and the Micro Trader. On one hand, it fulfils each tranche's trading task given by the Macro Trader. On the other hand, it adjusts order execution progress of the Micro Trader by issuing a subgoal appropriate for the short-term liquidity, where the subgoal works as a high-level action abstraction to simplify the Micro Trader's decision-making.

\begin{figure*}[!htb]
	\centering\includegraphics[width=0.85\linewidth]{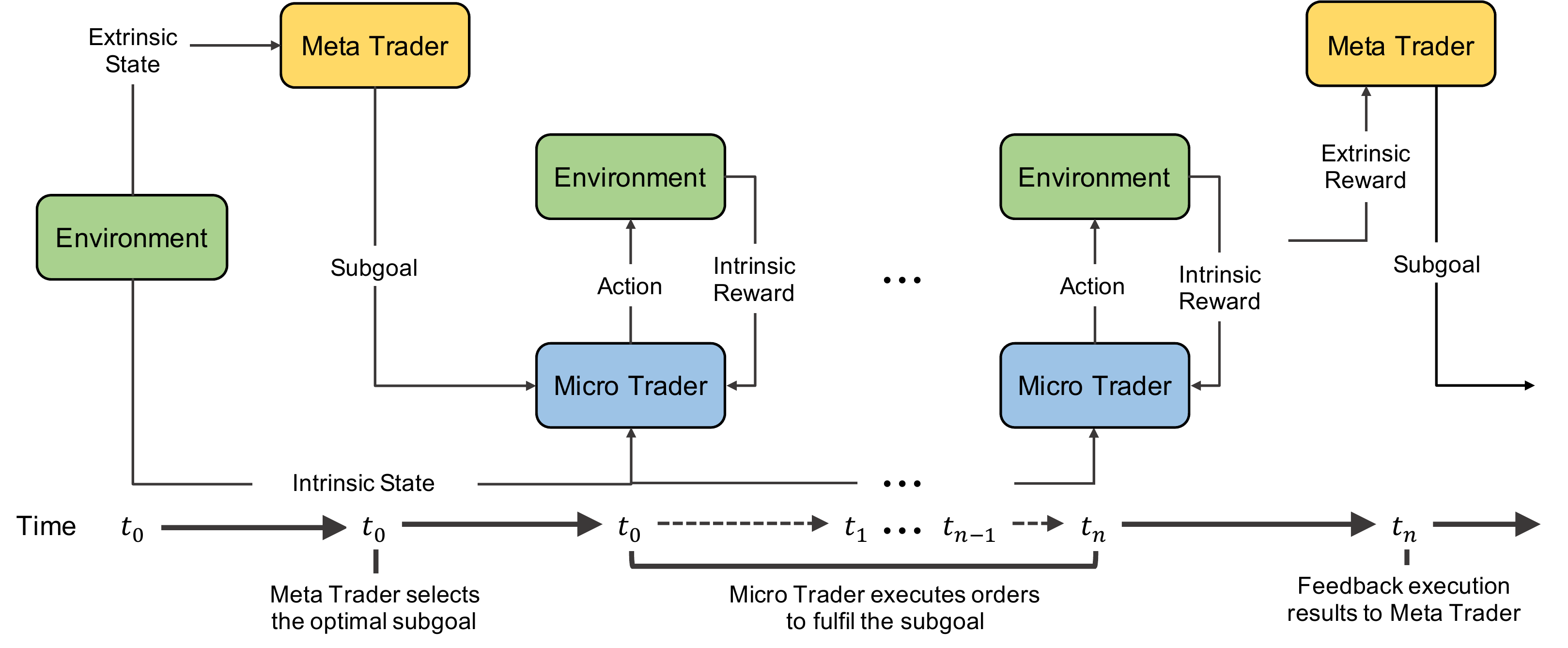}
	\caption{Workflow of Meta Trader and Micro Trader. The Meta Trader first specifies an optimal subgoal for the Micro Trader. Then the Micro Trader executes orders in the simulation trading environment to fulfil the subgoal. After that, the Meta Trader gets an extrinsic reward as the subgoal's execution feedback and specifies a new optimal subgoal for the Micro Trader.}
	\label{fig:meta-micro}
\end{figure*}

As Fig.~\ref{fig:meta-micro} shows, the Meta Trader first selects the subgoal $g^\ast$ with the highest expected reward from a preset subgoal set $\mathcal{G}$ based on current extrinsic state $\widetilde{s}$ and its policy $\pi_g$ to generate a mini-tranche. Assuming $\pi_g$ is a greedy policy, the optimal subgoal is selected by,
\begin{equation}
	\label{eq:subgoal-select}
	g^\ast=\left(T_{g^\ast},V_{g^\ast}\right)=\mathop{\arg\max}_{g\in\mathcal{G}}{Q(\widetilde{s},g;\theta)},
\end{equation}
where $Q(\cdot)$ is the state-action value function of Meta Trader estimated by a multilayer perception (MLP), $T$ is the maximum step and $V$ is the size of the order to be executed.

Then the Micro Trader receives $g^\ast$ and executes child orders to fulfil $g^\ast$. At the end of each subgoal execution, the unfilled orders will be issued as a market order. The VWAP slippage during this period is used as $\widetilde{r}$ and feeds back to the Meta Trader. We leverage DDQN~\cite{van2016deep} to iteratively update $Q(\cdot)$ and $\pi_g$. The DDQN consists of two independent state-action value functions $Q_2$ and $Q_2'$ with the same structure, where $Q_1$ is estimated by,

\begin{footnotesize}
	\begin{equation}
		\!Q_1(\! \widetilde{s}, g; \theta_1 \!) \!=\! \max_{\pi_g} \mathbb{E} \! \left[ \sum_{t'=t}^{t+N}\!\widetilde{r}_{t'} \!+\! \gamma Q_1'({\widetilde{s}}_{t \! + \! N}, \! g'; \! \theta_1') \Bigg|{\widetilde{s}}_t=\widetilde{s},  g_t \!=\! g,\pi_g \! \right]\!,
	\end{equation}
\end{footnotesize}%
where $N$ is actual execution step of $g$, and $g'$ is selected by Eq. \ref{eq:subgoal-select}. We use mean square error (MSE) as loss function to optimize $Q_1$'s and $Q_1'$'s parameters. Taking $Q_1$ as an example, its parameters $\theta_1$ are updated by,

\begin{footnotesize}
	\begin{equation}
		\begin{aligned}
			&\theta_{1|t+N}={\mathop{ \arg\min}_{\theta_{1|t}}{\frac{1}{\left|D_1\right|}}}\\
			&~~~\sum_{(\widetilde{s}, g, \widetilde{r}, \widetilde{s}')\sim\mathcal{D}_1} \! \left( \sum_{t'=t}^{t+N}\!\widetilde{r}_{t'} \! +\! \gamma{Q_1'\left({\widetilde{s}}',g';\theta_{1|t}'\right)} \!-\! Q_1\left({\widetilde{s}}_t,g_t;\theta_{1|t}\right)\!\right)^2\!,
		\end{aligned}
	\end{equation}
\end{footnotesize}%
where $\mathcal{D}_1$ is the experience replay of the Meta Trader consisting of $(\widetilde{s}, g, \widetilde{r}, \widetilde{s}')$, and $|D_1|$ is the batch size sampled from $\mathcal{D}_1$.

\subsection{Micro Trader}
\begin{figure}[htb]
	\centering
	\includegraphics[width=\linewidth]{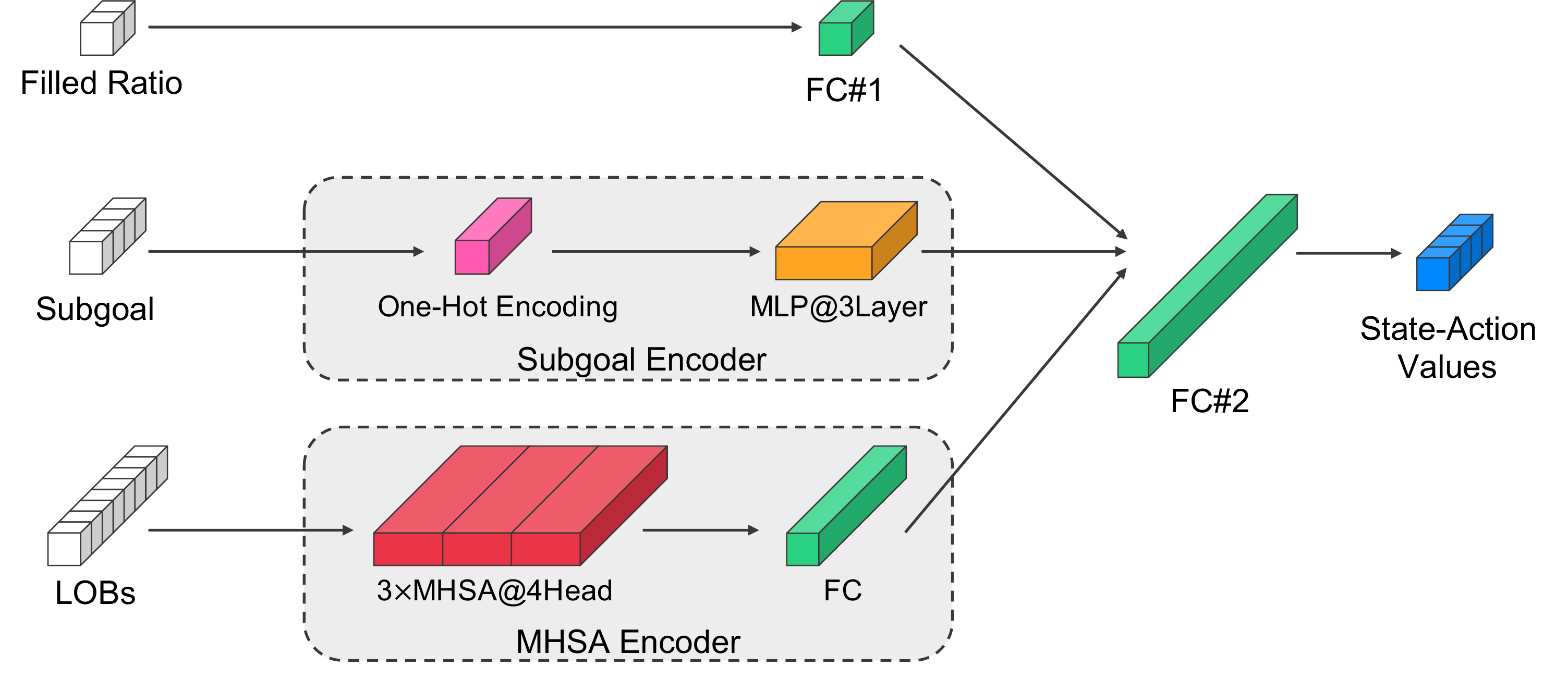}
	\caption{The structure of Micro Trader, where the MHSA encoder includes three layers four-headed self-attention encoder, and MLP consists of three FC layers.}
	\label{fig:micro-trader}
\end{figure}

As the lowest layer of the M3T, the Micro Trader is used to execute each trading step. It optimizes order execution and finally fulfils the subgoal given by the Meta Trader. As shown in Fig.~\ref{fig:micro-trader}, the Micro Trader includes an multi-headed self-attention (MHSA) encoder (\emph{a.k.a.} Transformer encoder~\cite{vaswani2017attention}), an FC layer, and a subgoal encoder to process different sources of data. The MHSA encoder extracts the temporal representation of markets from the last 20 LOB snapshots. The FC layer maps the execution ratio into hidden representation. The subgoal encoder first uses one-hot encoding to encode the subgoal given by the Meta Trader and then learns a representation of the subgoal by an MLP. After that, another FC layer connects the above three modules and maps their outputs to state-action values.

The Micro Trader extends the general RL policy to include the subgoal given by the Meta Trader. At time $t$, it receives subgoal $g$ given by the Meta Trader and decides order execution according to its policy $\pi_{ag}$. For instance, assuming $\pi_{ag}$ is a greedy strategy, the order is executed by,
\begin{equation}
	\label{eq:action-select}
	a^\ast=\mathop{\arg\max}_ {a\in\mathcal{A}}{Q\left(s,a;g,\theta\right)}.
\end{equation}

Similarly, the Micro Trader uses the DDQN to optimize its order execution policy and fulfil subgoals. The two independent state-action value functions $Q_2$ and $Q_2'$ of the Micro Trader is extended to include subgoals, where $Q_2$ is estimated by,

\begin{footnotesize}
	\begin{equation}
		Q_2(\!s,\!a;\!g,\!\theta_2) \!=\! \max_{\pi_{ag}} \!{\mathbb{E} \!\left[r_t \!+\! \gamma{Q_2'\left(\!s'\!,\!a'\!;\!g\!,\theta_2'\right)}\Bigg|s_t\!=\!s,a_t\!=\!a,g_t\!=\!g,\pi_{ag}\!\right]},
	\end{equation}
\end{footnotesize}%
where $a'$ is selected by Eq.~\ref{eq:action-select}. We also use MSE as the loss function to optimize $Q_2$'s and $Q_2'$'s parameters. Taking $Q_2$ as an example, its parameters $\theta_2$ are updated by,

\begin{footnotesize}
	\begin{equation}
		\begin{aligned}
			&\theta_{2|t+1}={\mathop{\arg\min}_{\theta_{2|t}}{\frac{1}{|D_2|}}}\\
			&~~\sum_{(s,a,r,s\!'\!,g)\sim\mathcal{D}_2}\!\left(\!r_t\!+\!\gamma\max_{a\!'}{Q_2'\left(s\!',a\!';g,\theta_{2|t}\!'\right)}\!-\!Q_2\left(s_t,a_t;g,\theta_{2|t}\right)\!\right)^2\!,
		\end{aligned}
	\end{equation}
\end{footnotesize}%
where $\mathcal{D}_2$ is the experience replay of the Micro Trader consisting of $(s, a, g, r, s')$, and $|D_2|$ is the batch size sampled from $\mathcal{D}_2$.

Algorithm~\ref{alg:m3t} summarizes the process of executing a parent order $\mathcal{T}$ via the M3T, which reduces transaction costs by executing orders hierarchically.

\begin{algorithm}[htb]  
	\caption{Parent Order Execution via M3T}  
	\label{alg:m3t}  
	\begin{algorithmic}[1]  
		\REQUIRE historical volume profile $\{vp_\tau^{(i)}\}_{\tau=t-20}^{t-1}$; parent order $\mathcal{T}$, Macro Trader $\mathcal{M}_{Macro}$, Meta Trader $\mathcal{M}_{Meta}$; Micro Trader $\mathcal{M}_{Micro}$; trading environment $\mathcal{E}_t$.
		\STATE $\mathcal{M}_{Macro}$ estimates $vp_t^{(i)}$ based on $\{vp_\tau^{(i)}\}_{\tau=t-20}^{t-1}$;
		\STATE $\mathcal{M}_{Macro}$ allocates $\mathcal{T}$ into tranches; $\{\mathcal{T}_t^{(i)}\}_{i=1}^8$
		\FOR {$\mathcal{T}_t^{(i)}$ $\{\mathcal{T}_t^{ (i)}\}_{i=1}^8$}
		\WHILE{$\mathcal{T}_t^{(i)}$ is unfilled}
		\STATE $\mathcal{M}_{Meta}$ selects a subgoal $g$ according to $\widetilde{s}$;
		\WHILE{$g$ is unfilled}
		\STATE $\mathcal{M}_{Micro}$ executes orders in $\mathcal{E}_t$;
		\ENDWHILE
		\ENDWHILE
		\ENDFOR
	\end{algorithmic}
\end{algorithm}

\section{Experiments}
\label{sec:exp}
Our experiments evaluate the transaction costs of different algorithmic trading methods, volume profile estimation methods, as well as DL and RL backbones by back testing.

\subsection{Dataset}
From the SSE 50 Index's constituent stocks in May 2014, we select eight representative stocks from different sectors and list them in Table \ref{tab:stock}. We perform experiments on their level-2 data (consist of LOBs and trade) for each trading day from June 2014 to March 2015. The LOBs are snapshots of quotes with about 6M overall, while the number of trading data is about 500M. We exclude the data of the days when the market price reaches the price limit due to illiquidity.

\begin{table}[htb]
	\caption{The ID and Name of the Stocks in Our Experiments.}
	\centering
	\begin{tabular}{|p{40pt}|p{175pt}|}
		\hline
		\textbf{Stock ID} & \textbf{Stock name}\\
		\hline
		600000	&	Shanghai Pudong Development Bank\\
		600010	&	Inner Mongolia Baotou Steel Union\\
		600018	&	Shanghai International Port\\
		600028	&	China Petroleum \& Chemical Corporation\\
		600030	&	Citic Securities Company Limited\\
		600048	&	Poly Developments and Holdings Group\\
		600050	&	China United Network Communications	\\
		600104	&	Saic Motor	\\
		\hline
	\end{tabular}
	\label{tab:stock}
\end{table}

\subsection{Baselines and Metrics}
To verify the effectiveness of the Macro Trader in volume profile estimation, we compare it with a classical volume profile estimation method and two DL methods:
\begin{itemize}
	\item \textbf{MA}: Use the average value of the last $N$ days ( 20 days in our experiments) as the estimate of the future.
	\item \textbf{FC}: A neural network consisting of a single layer of learnable parameters, \emph{a.k.a.} linear layer.
	\item \textbf{MLP}: A neural network consisting of several layers of learnable parameters.
\end{itemize}

We compare our M3T with two classical algorithmic trading strategies and three deep RL methods:
\begin{itemize}
	\item \textbf{VWAP}: The order is allocated into tranches according to volume profiles to track the VWAP.
	\item \textbf{Arrival Price (AP)}: Order's price is determined by the filled ratio, \emph{i.e.,} if the filled ratio is lower than the time ratio, execute as a market order; otherwise, execute as a limit order. When the filled ratio exceeds 10\% of the time ratio, it stops issuing new orders.
	\item \textbf{DQN}~\cite{mnih2015human}: A TD-based deep RL model. \cite{nevmyvaka2006reinforcement}\cite{almgren2001optimal}\cite{shen2014risk} used it to optimize algorithmic trading strategies.
	\item \textbf{DDQN}~\cite{van2016deep}: An improved deep RL model based on DQN. Ning~\emph{et al.}~\cite{ning2018double} used it to optimize algorithmic trading strategies.
	\item \textbf{Advantage Actor-Critic (A2C)}~\cite{wu2017openai}: An actor-critic-based deep RL model, which consists of an actor network and a critic network. 
\end{itemize}%
The classical algorithmic trading strategies use historical volume profiles' MA to estimate future volume profiles, while the DRL methods use LSTM. Table~\ref{tab:baseline} summarizes the settings of these baselines and the M3T in our experiments.

\begin{table}[htb]
	\caption{Structure Setting of Baselines and M3T.\label{tab:baseline}}
	\begin{tabular}{|l|l|l|l|}
		\hline
		\multirow{2}{*}{\textbf{Method}} &  \textbf{Macro Trader}	&	\textbf{Meta Trader}	&	\textbf{Micro Trader}\\
		&	\scriptsize\textbf{(Profile Estimation)}	&	\scriptsize\textbf{(Subgoal Selection)}	&	\scriptsize\textbf{(Order Execution)}\\
		\hline
		VWAP & MA & N/A & Rule-based \\
		AP & MA & N/A & Rule-based \\
		DQN & LSTM & N/A & DQN@MHSA \\
		DDQN & LSTM & N/A & DDQN@MHSA \\
		A2C & LSTM & N/A & A2C@MHSA \\
		M3T & LSTM & DDQN@MLP & DDQN@MHSA \\
		\hline
	\end{tabular}
\end{table}

We use the VWAP slippage presented in Eq. \ref{eq:slippage} to measure the transaction costs and propose \textit{subgoal speed} to evaluate the overall liquidity required by subgoals within each tranche, that is,
\begin{equation}
	u=\frac{\sum_{g\in {\text{tranche}_i}}\left(\left(V_g/T_g\right)\cdot\ n_g\right)}{\sum_{g\in {\text{tranche}_i}} n_g},
\end{equation}
where $n_g$ is the times that Meta Trader selects subgoal $g$ within tranche $\#i$.

\subsection{Training Setup}
In Macro Trader training, we use the first 60\% of the dataset (June 2014 to October 2014) as the training set, the middle 20\% of the dataset (November 2014 to December 2014) as the validation set, and the last 20\% of the dataset (January 2015 to March 2015) as the test set. In Meta Trader and Micro Trader training, we use the first 80\% of the dataset (June 2014 to December 2014) as the training set and the last 20\% of the dataset (January 2015 to March 2015) as the test set. 

We train the M3T, DQN, and DDQN with $\epsilon$-greedy and experience replay to improve these models' performance. The M3T and the baselines are trained by the same hyperparameters selected according to the performance of DDQN on \textit{Shanghai Pudong Development Bank (600000)}, as shown in Table~\ref{tab:hyper-param}.

\begin{table}[htb]
	\caption{Hyperparameters of M3T and Baselines.\label{tab:hyper-param}}
	\begin{tabular}{|p{50pt}|p{100pt}|p{65pt}|}
		\hline
		\textbf{Method} &  \textbf{Hyperparameter} & \textbf{Best/Default}\\
		\hline
		\multirow{3}{*}{Macro Trader} & Epoch & 5e3 \\
		& Batch size & All samples \\
		& Learning rate & 1e-4 \\
		\hline
		\multirow{4}{*}{RL Training} & Episode &  1e4 \\
		& Discount  & 0.99 \\
		& Learning rate  & 5e-5 \\
		& Tranche size  & [100K,  200K] \\
		\hline
		Meta Trader & Initial $\epsilon$  & 1.00\\
		Micro Trader & $\epsilon$ decline rate  & $0.99 / 5$ Episodes\\
		DQN & Experience replay capacity  & 1e4 \\
		DDQN & Experience replay batch size  & 128 \\
		\hline
		A2C & Synchronized environments  & 16 \\
		\hline
	\end{tabular}
\end{table}

\subsection{Experimental Results}
\subsubsection{Volume Profile Estimation}
To reveal the performance of macro traders in volume profile estimation, we perform experiments on the test set and calculate the MSE of the estimated volume profiles. Table~\ref{tab:macro} shows the MSE of the volume profile estimation over eight stocks, where the unit of values is millesimal (\emph{i.e.}, 1e-3). Each stock's best and second-best results are highlighted in bold and underlined. The experimental results reveal that the Macro Trader has the lowest MSE over all stocks, confirming its effectiveness in volume profile estimation. MA shows the second-best performance and far outperforms FC and MLP.

\begin{table*}[htb]
	\caption{MSE (1e-3) of Macro Trader and Baselines in Volume Profile Estimation.\label{tab:macro}}
	\centering
	\begin{tabular}{|p{50pt}|p{35pt}|p{35pt}|p{35pt}|p{35pt}|p{35pt}|p{35pt}|p{35pt}|p{35pt}|}
		\hline
		\textbf{Model}	&	\textbf{600000}		&	\textbf{600010}		&	\textbf{600018}	&	\textbf{600028}	&	\textbf{600030}	& \textbf{600048}	&	\textbf{600050} & \textbf{600104} \\
		\hline
		MA	&	\underline{2.48}	&	\underline{3.41}	&	\underline{4.43}	&	\underline{2.18}	&	\underline{3.05}	&	\underline{2.32}	&	\underline{2.71}	&	\underline{1.64}	\\
		Linear	&	3.09	&	3.40	&	4.59	&	2.61	&	4.41	&	2.63	&	3.49	&	1.79	\\
		MLP	&	4.66	&	6.02	&	5.86	&	4.22	&	6.85	&	5.31	&	4.62	&	2.92	\\
		\hline
		Macro Trader	&	\textbf{2.27}	&	\textbf{3.17}	&	\textbf{4.20}	&	\textbf{2.16}	&	\textbf{2.96}	&	\textbf{2.06}	&	\textbf{2.68}	&	\textbf{1.62}	\\
		\hline
	\end{tabular}
\end{table*}

\subsubsection{Transaction Costs}
We evaluate the transaction costs of M3T and baselines by simulating trading 120,000 shares for each day. Table~\ref{tab:vwap} compares transaction costs of the M3T and the baselines on eight stocks in terms of VWAP slippage (bp). The results show that the M3T outperforms all the baselines and achieves the lowest transaction costs over seven stocks. DDQN is the best baseline, achieving second-best over four stocks, while DQN achieves second-best over two stocks. The AP's general performance is close to the DQN and slightly better than the A2C. The VWAP strategy has the highest transaction costs. These results demonstrate that using RL models to optimize the VWAP strategy can reduce transaction costs. Furthermore, even though the DDQN adapts the same RL backbone method as the M3T, the M3T's transaction costs is lower than the DDQN by an average of 1.16 bp and by 2.22 bp at most, demonstrating that the M3T can further reduce the transaction costs than existing methods.

\begin{table*}[htb]
	\caption{VWAP Slippage (bp) of M3T and Baselines in Parent Order Execution.\label{tab:vwap}}
	\centering
	\begin{tabular}{|p{50pt}|p{43pt}|p{43pt}|p{43pt}|p{43pt}|p{43pt}|p{43pt}|p{43pt}|p{43pt}|}
		\hline
		\textbf{Model}	&	\textbf{600000}		&	\textbf{600010}		&	\textbf{600018}	&	\textbf{600028}	&	\textbf{600030}	& \textbf{600048}	&	\textbf{600050} & \textbf{600104} \\
		\hline
		VWAP	&	-3.37\scriptsize$\pm$4.25	&	-10.47\scriptsize$\pm$8.77	&	-9.01\scriptsize$\pm$6.59	&	-8.21\scriptsize$\pm$6.05	&	-2.97\scriptsize$\pm$6.31	&	-6.64\scriptsize$\pm$5.88	&	-11.38\scriptsize$\pm$12.12	&	-3.23\scriptsize$\pm$5.80	\\
		AP	&	~2.41\scriptsize$\pm$4.65	&	~~~5.37\scriptsize$\pm$7.97	&	-1.58\scriptsize$\pm$6.33	&	~\textbf{5.77}\scriptsize$\pm$7.02	&	~0.51\scriptsize$\pm$4.18	&	~1.57\scriptsize$\pm$6.06	&	~~~\underline{2.41}\scriptsize$\pm$10.89	&	~0.10\scriptsize$\pm$5.34	\\
		\hline
		DQN	&	~3.16\scriptsize$\pm$3.73	&	~10.64\scriptsize$\pm$6.59	&	~6.05\scriptsize$\pm$3.58	&	~3.01\scriptsize$\pm$5.12	&	~\underline{2.33}\scriptsize$\pm$5.49	&	~1.92\scriptsize$\pm$6.73	&	~~-4.89\scriptsize$\pm$9.85	&	~\underline{3.00}\scriptsize$\pm$6.21	\\
		DDQN	&	~\underline{3.53}\scriptsize$\pm$3.99	&	~\underline{10.92}\scriptsize$\pm$6.22	&	~\underline{7.60}\scriptsize$\pm$5.98	&	~4.51\scriptsize$\pm$6.78	&	~2.32\scriptsize$\pm$4.97	&	~\underline{2.31}\scriptsize$\pm$4.63	&	~~~0.33\scriptsize$\pm$6.63	&	~2.98\scriptsize$\pm$7.00	\\
		A2C	&	~2.35\scriptsize$\pm$3.95	&	~~~4.75\scriptsize$\pm$8.03	&	~3.58\scriptsize$\pm$9.28	&	~3.23\scriptsize$\pm$5.88	&	~1.97\scriptsize$\pm$4.22	&	-0.85\scriptsize$\pm$5.73	&	~~-2.42\scriptsize$\pm$8.03	&	~2.38\scriptsize$\pm$4.54	\\
		\hline
		M3T	&	~\textbf{4.38}\scriptsize$\pm$3.76	&	~\textbf{12.18}\scriptsize$\pm$7.00	&	~\textbf{9.42}\scriptsize$\pm$6.12	&	~\underline{5.23}\scriptsize$\pm$4.89	&	~\textbf{3.01}\scriptsize$\pm$5.28	&	~\textbf{3.32}\scriptsize$\pm$6.11	&	~~~\textbf{2.55}\scriptsize$\pm$8.76	&	~\textbf{3.66}\scriptsize$\pm$5.45	\\
		\hline
	\end{tabular}
\end{table*}

\begin{table*}[htb]
	\caption{VWAP Slippage (bp) of Different Volume Profile Estimation Methods.\label{tab:vp}}
	\centering
	\begin{tabular}{|p{50pt}|p{35pt}|p{35pt}|p{35pt}|p{35pt}|p{35pt}|p{35pt}|p{35pt}|p{35pt}|}
		\hline
		\textbf{Model}	&	\textbf{600000}		&	\textbf{600010}		&	\textbf{600018}	&	\textbf{600028}	&	\textbf{600030}	& \textbf{600048}	&	\textbf{600050} & \textbf{600104} \\
		\hline
		VWAP@MA	&	-3.37	&	-10.47	&	-9.01	&	-8.21	&	-2.97	&	-6.64	&	-11.38	&	-3.23	\\
		VWAP@LSTM	&	-3.35	&	-10.51	&	-8.92	&	-8.21	&	-2.97	&	-6.63	&	-11.37	&	-3.20	\\
		$\Delta_{(\text{LSTM}-\text{MA})}$	&	\textbf{+0.02}	&	\underline{-0.04}	&	\textbf{+0.09}	&	\underline{0.00}	&	\underline{0.00}	&	\textbf{+0.01}	&	\textbf{+0.01}	&	\textbf{+0.03}	\\
		\hline
		AP@MA	&	2.41	&	5.37	&	-1.58	&	5.77	&	0.51	&	1.57	&	2.41	&	0.10	\\
		AP@LSTM	&	2.42	&	5.38	&	-1.62	&	5.80	&	0.50	&	1.60	&	2.38	&	0.06	\\
		$\Delta_{(\text{LSTM}-\text{MA})}$	&	\textbf{+0.01}	&	\underline{-0.01}	&	\underline{-0.04}	&	\textbf{+0.03}	&	\underline{-0.01}	&	\textbf{+0.03}	&	\underline{-0.03}	&	\underline{-0.04}	\\
		\hline
		DQN@MA	&	3.12	&	10.17	&	6.02	&	2.89	&	2.12	&	1.77	&	-5.28	&	2.56	\\
		DQN@LSTM	&	3.16	&	10.64	&	6.05	&	3.01	&	2.33	&	1.92	&	-4.89	&	3.00	\\
		$\Delta_{(\text{LSTM}-\text{MA})}$	&	\textbf{+0.04}	&	\textbf{+0.47}	&	\textbf{+0.03}	&	\textbf{+0.12}	&	\textbf{+0.21}	&	\textbf{+0.15}	&	\textbf{+0.39}	&	\textbf{+0.44}	\\
		\hline
		DDQN@MA	&	3.47	&	10.83	&	7.44	&	4.50	&	2.33	&	2.19	&	0.29	&	2.73	\\
		DDQN@LSTM	&	3.53	&	10.92	&	7.60	&	4.51	&	2.32	&	2.31	&	0.33	&	2.98	\\
		$\Delta_{(\text{LSTM}-\text{MA})}$	&\textbf{+0.06}&	\textbf{+0.09}	&	\textbf{+0.16}	&	\textbf{+0.01}	&	\underline{-0.01}	&	\textbf{+0.12}	&	\textbf{+0.04}	&	\textbf{+0.25}	\\
		\hline
		A2C@MA	&	2.33	&	4.70	&	3.47	&	3.21	&	1.99	&	-0.92	&	-2.47	&	2.39	\\
		A2C@LSTM	&	2.35	&	4.75	&	3.58	&	3.23	&	1.97	&	-0.85	&	-2.42	&	2.38	\\
		$\Delta_{(\text{LSTM}-\text{MA})}$	&	\textbf{+0.02}	&	\textbf{+0.05}	&	\textbf{+0.11}	&	\textbf{+0.02}	&	\underline{-0.02}	&	\textbf{+0.07}	&	\textbf{+0.05}	&	\underline{-0.01}	\\
		\hline
		M3T@MA	&	4.35	&	11.09	&	9.39	&	5.18	&	3.03	&	3.09	&	2.32	&	3.51	\\
		M3T@LSTM	&	4.38	&	12.18	&	9.42	&	5.23	&	3.01	&	3.32	&	2.55	&	3.66	\\
		$\Delta_{(\text{LSTM}-\text{MA})}$ &	\textbf{+0.03}	&	\textbf{+0.09}	&	\textbf{+0.03}	&	\textbf{+0.05}	&	\underline{-0.02}	&	\textbf{+0.13}	&	\textbf{+0.23}	&	\textbf{+0.15}	\\
		\hline
	\end{tabular}
\end{table*}

\begin{table*}[htb]
	\caption{VWAP Slippage (bp) of M3T Based on Different DL Models.\label{tab:dl}}
	\centering
	\begin{tabular}{|p{50pt}|p{35pt}|p{35pt}|p{35pt}|p{35pt}|p{35pt}|p{35pt}|p{35pt}|p{35pt}|}
		\hline
		\textbf{Model}	&	\textbf{600000}		&	\textbf{600010}		&	\textbf{600018}	&	\textbf{600028}	&	\textbf{600030}	& \textbf{600048}	&	\textbf{600050} & \textbf{600104} \\
		\hline
		M3T@FC	&	3.21	&	-85.12	&	6.29	&	-8.72	&	0.88	&	0.73	&	-38.24	&	0.03	\\
		M3T@CNN	&	-3.58	&	-13.45	&	-2.15	&	-7.23	&	-0.50	&	-2.12	&	-28.45	&	-1.75	\\
		M3T@LSTM	&	\underline{4.12}	&	\underline{11.22}	&	\underline{8.63}	&	\underline{4.25}	&	\underline{2.33}	&	\textbf{3.47}	&	\underline{1.53}	&	\underline{3.47}	\\
		M3T@MHSA	&	\textbf{4.38}	&	\textbf{12.18}	&	\textbf{9.42}	&	\textbf{5.23}	&	\textbf{3.01}	&	3.32	&	\textbf{2.55}	&	\textbf{3.66}	\\
		\hline
	\end{tabular}
\end{table*}

\begin{table*}[htb]
	\caption{VWAP Slippage (bp) of M3T Based on Different RL Models.\label{tab:rl}}
	\centering
	\begin{tabular}{|p{50pt}|p{35pt}|p{35pt}|p{35pt}|p{35pt}|p{35pt}|p{35pt}|p{35pt}|p{35pt}|}
		\hline
		\textbf{Model} & \textbf{600000} & \textbf{600010} & \textbf{600018} & \textbf{600028} & \textbf{600030} & \textbf{600048} & \textbf{600050} & \textbf{600104} \\ \hline
		M3T@DQN        & 3.45            & 10.95           & 8.83            & \textbf{5.51}   & 2.97            & 3.21            & 2.25            & 3.39            \\
		M3T@DDQN       & \textbf{4.38}   & \textbf{12.18}  & \textbf{9.42}   & 5.23            & \textbf{3.01}   & \textbf{3.32}   & \textbf{2.55}   & \textbf{3.66}   \\ \hline
	\end{tabular}
\end{table*}

\subsubsection{Computing Costs}
Since market opportunities are fleeting, algorithmic trading's time efficiency is a critical indicator in practice. To reveal the time efficiency of M3T, we measure the time costs of training M3T and baselines on an Nvidia Titan XP GPU. Table~\ref{tab:macro-time} shows the average time cost of training Macro Trader is less than 10 seconds per 1000 epochs, which is acceptable for users because the Macro Trader only needs to train offline once before executing orders. Table~\ref{tab:drl-time} shows that even though the M3T includes an additional hierarchical trader (i.e., the Meta Trader) compared to the DDQN, the average cost for training M3T is similar to the DQN and the DDQN used in previous studies, and it is lower than the A2C.
\begin{table}[htb]
	\caption{Time Costs (per 1000 epochs) of Training Macro Trader.\label{tab:macro-time}}
	\centering
		\begin{tabular}{|p{55pt}|p{42pt}|p{42pt}|p{42pt}|}
			\hline
			\textbf{Macro Trader}	&	\textbf{Linear}		&	\textbf{MLP}		&	\textbf{LSTM}\\
			\hline
			\textbf{Time costs}	&	1.23s\scriptsize{$\pm$0.22}	&	10.65s\scriptsize{$\pm$0.95}	&	7.44s\scriptsize{$\pm$0.77}\\
			\hline
	\end{tabular}
\end{table}

\begin{table}[htb]
	\caption{Time Costs (per episode) of Training M3T and Baselines.\label{tab:drl-time}}
	\centering
		\begin{tabular}{|p{39pt}|p{38pt}|p{38pt}|p{42pt}|p{38pt}|}
			\hline
			\textbf{Methods}	&	\textbf{DQN}		&	\textbf{DDQN}	&	\textbf{A2C}	&	\textbf{M3T}\\
			\hline
			\textbf{Time costs}	&	81.75s\scriptsize{$\pm$19.88}&	87.08s\scriptsize{$\pm$16.34}&	126.59s\scriptsize{$\pm$18.21}	& 98.42s\scriptsize{$\pm$21.06}\\
			\hline
	\end{tabular}
\end{table}

\subsection{Analysis}
Although we have demonstrated the effectiveness of M3T in reducing transaction costs via previous experiments, we would like to figure out which module in the M3T effects on reducing the transaction costs. In other words, we are concerned with the following three questions:
\begin{enumerate}
\item Can accurate prediction of volume profiles reduce transaction costs?

\item How do different DL backbones affect M3T’s performance?

\item How do different RL backbones affect M3T’s performance?

\end{enumerate}

To answer the above three questions empirically, we added the following three analytical experiments:

\textbf{Volume Profile Estimation}:
In order to explore the impact of different volume profile estimation methods on the transaction costs, we respectively use MA and LSTM to estimate the volume profiles and compare the final transaction costs of the M3T and the baselines, as shown in Table~\ref{tab:vp}. The value highlighted in bold indicates that the final transaction costs of using LSTM are lower than using MA, while the value highlighted in underline is the opposite. The results show that using LSTM can reduce transaction costs. Referring to the results in Table~\ref{tab:macro}, we can conclude that accurately estimating volume profiles effectively reduces transaction costs.

\textbf{Deep Learning Backbone}:
To reveal the impact of different DL models on the M3T's performance, we establish Micro Traders based on several DL models such as FC, CNN, and LSTM as comparisons. The results in Table~\ref{tab:dl} demonstrate that the MHSA has the best performance, the LSTM is second, and the CNN performs the worst.

\textbf{Reinforcement Learning Backbone}:
To reveal the impact of different RL models on the M3T's performance, we train a Meta Trader and a Micro Trader based on DQN as comparisons. The results shown in Table~\ref{tab:rl} demonstrate that DDQN-based M3T outperforms DQN on almost all stocks.

\begin{figure*}[!h]
	\begin{minipage}{\linewidth}
		\centering 
		\includegraphics[width=0.96\linewidth]{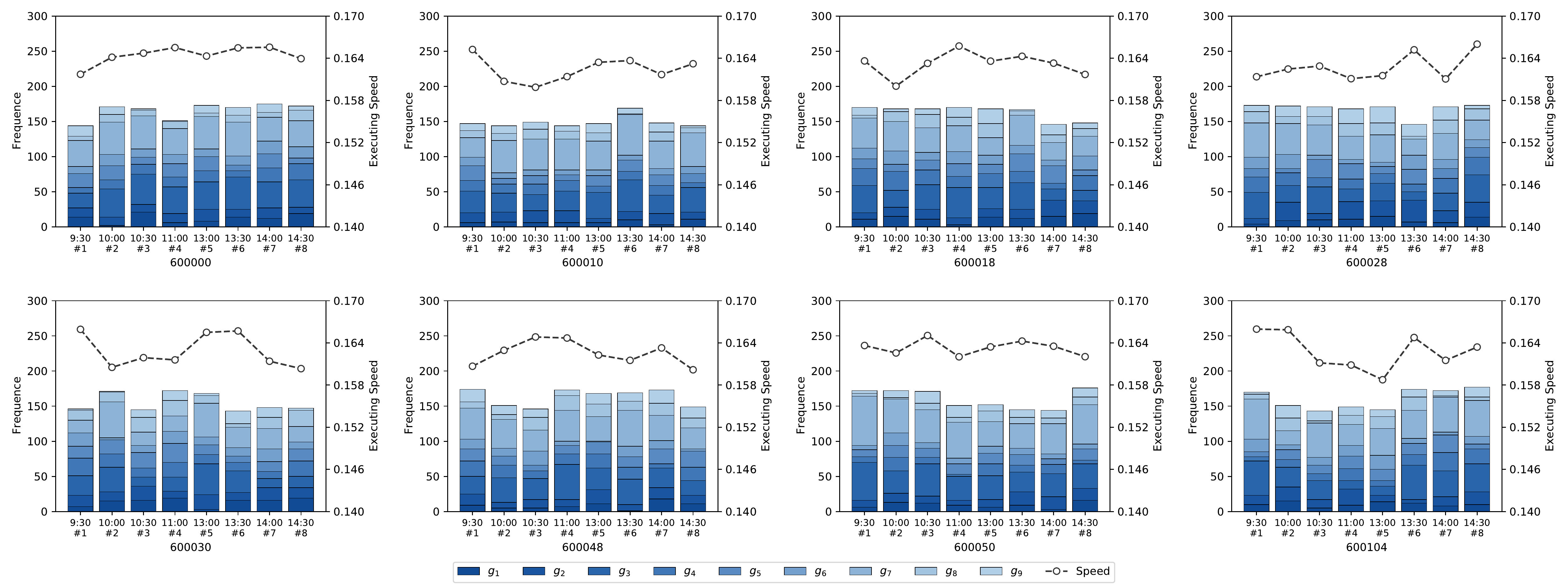}\\
		\footnotesize (a) M3T@DDQN
	\end{minipage}
	\vspace{5pt}
	
	\begin{minipage}{\linewidth}
		\centering 
		\includegraphics[width=0.96\linewidth]{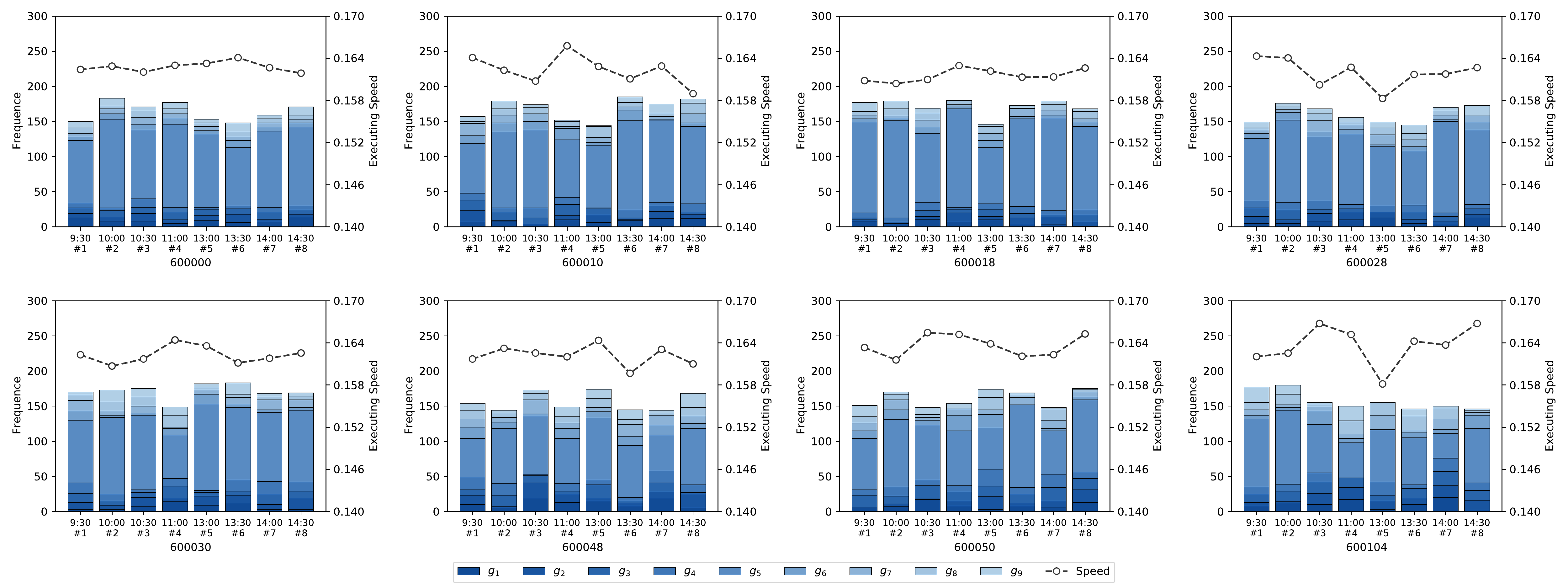}\\
		\footnotesize (b) M3T@DQN
	\end{minipage}
	\caption{Subgoal counts and subgoal speed over eight stocks.\label{fig:subgoal}}
\end{figure*}

\subsection{Case Study}
In order to analyze whether the Meta Trader can select appropriate subgoals according to the market's liquidity, we count the subgoals selected by the Meta Trader of the best-performed M3T (i.e., the M3T with the lowest transaction cost over checkpoints) in each tranche and calculate their \textit{subgoal speed}, as shown in  Fig.~\ref{fig:subgoal}. The curve in Fig.~\ref{fig:subgoal} shows that the \textit{subgoal speed} of the M3T@DQN and the M3T@DDQN is generally between 0.16 and 0.17. It performs the same trend on most stocks, with higher \textit{subgoal speed} in the middle and lower on both sides, indicating that the M3T@DQN and the M3T@DDQN tend to fulfil orders quickly in the less liquid period. By comparing the number of subgoals selected by the M3T@DQN and the M3T@DDQN, we find that the M3T@DQN tends to select the subgoals with balanced \textit{subgoal speed}, such as subgoal $\#5$, while the M3T@DDQN tends to select the subgoals with relatively slower (\emph{e.g.}, subgoal $\#3$) or faster \textit{subgoal speed} (\emph{e.g.}, subgoal $\#7$). Combining with the results shown in Table~\ref{tab:vwap}, it demonstrates that the Meta Trader can reduce transaction costs by adjusting order execution speed to keep consistency with the short-term liquidity.

\section{Conclusions}
\label{sec:con}
In this paper, we discuss a critical concern of brokers in designing an intelligent volume-weighted average price (VWAP) strategy to achieve a lower transaction cost in a dynamic market. Although many studies have demonstrated that using RL to execute orders can reduce transaction costs, these methods only apply to algorithmic trading tasks with short duration and small order sizes. To address the shortcomings of existing RL methods in executing the VWAP strategy, we propose a deep learning and HRL jointed architecture termed M3T. It consists of three traders that capture market patterns and execute orders from different temporal scales, improving VWAP strategy from tranche allocation, subgoal selection, and order execution. Experimental results demonstrate that 1) the M3T outperforms the baselines, with an average cost saving of 1.16 base points compared to the optimal baseline DDQN; 2) the Macro Trader can improve the accuracy of volume profile estimation and reduce transaction costs; 3) each trader in the M3T is effective to reduce the transaction costs. The additional subgoal selection analysis demonstrates that the Meta Trader can reduce transaction costs by selecting appropriate subgoals according to market liquidity.

In future work, we plan to quantify the impact of news events on the market as an additional market state. Besides, incorporating the neural architecture search to optimize M3T's hyperparameters may further improve its performance.

\section*{Acknowledgments}
This work was supported in part by the National Natural Science Foundation of China under Grant 61602149, in part by the Central University Basic Research Fund of China under Grant B210202078, and in part by the Postgraduate Research \& Practice Innovation Program of Jiangsu Province under Grant KYCX21\_0483.

\ifCLASSOPTIONcaptionsoff
  \newpage
\fi

\bibliographystyle{IEEEtran}
\bibliography{reference}




\end{document}